\documentclass[aps,pra,twocolumn,superscriptaddress]{revtex4}
\usepackage{graphicx}
\usepackage{amssymb}
\usepackage{amsmath}
\usepackage{bm}

\begin{document}

\title{Critical spin superflow in a spinor Bose-Einstein condensate}

\author{Joon Hyun Kim}
\affiliation{Department of Physics and Astronomy, and Institute of Applied Physics, Seoul National University, Seoul 08826, Korea}

\author{Sang Won Seo}
\affiliation{Department of Physics and Astronomy, and Institute of Applied Physics, Seoul National University, Seoul 08826, Korea}
\affiliation{Center for Correlated Electron Systems, Institute for Basic Science, Seoul 08826, Korea}

\author{Y. Shin}\email{yishin@snu.ac.kr}
\affiliation{Department of Physics and Astronomy, and Institute of Applied Physics, Seoul National University, Seoul 08826, Korea}
\affiliation{Center for Correlated Electron Systems, Institute for Basic Science, Seoul 08826, Korea}

%\date{\today}

\begin{abstract}
We investigate the critical dynamics of spin superflow in an easy-plane antiferromagnetic spinor Bose-Einstein condensate. Spin-dipole oscillations are induced in a trapped condensate by applying a linear magnetic field gradient and we observe that the damping rate increases rapidly as the field gradient increases above a certain critical value. The onset of dissipation is found to be associated with the generation of dark-bright solitons due to the modulation instability of the counterflow of two spin components. Spin turbulence emerges as the solitons decay because of their snake instability. We identify another critical point for spin superflow, in which transverse magnon excitations are dynamically generated via spin-exchanging collisions, which leads to the transient formation of axial polar spin domains.
\end{abstract}

\maketitle

The hallmark of a superfluid is its ability to flow without friction but the superflow becomes unstable above a certain critical velocity, generating excitations~\cite{Landau}. Detection of a finite critical velocity can serve as evidence of superfluidity and revealing its corresponding dissipation mechanism will improve our understanding of superfluidity. The questions regarding superflow stability become more interesting when a superfluid has internal spin degrees of freedom~\cite{Sonin10}. A superfluid has multiple superflow channels for mass and spin, and their interplay may allow new types of topological excitations, such as fractionalized quantum voritces~\cite{Jang11,Manni12,Seo15,Autti16} and spin-textured solitons~\cite{Bartolac81,Choi12,Hivet12}, possibly leading to qualitatively different dissipation dynamics~\cite{Choi13,Seo16,Ollikainen17}. Furthermore, the mixed mass and spin characters might give rise to ambiguities in the unique  determination of a critical velocity, particularly, when a system has spin-orbit coupling~\cite{Zhu12,Ozawa13}.

Critical spin superflow phenomena were first explored in experiments using superfluid helium-3, with the observation of phase slippage in a spin current~\cite{Borovik-Ramonov89}, and recently using two-component atomic Bose-Einstein condensate (BEC) systems, which can be regarded as effective spin-$1/2$ superfluids, demonstrating counterflow instability~\cite{Hamner11,Hoefer11, Beattie13}. A natural direction in which to extend the study is spinor BECs of spin-1 atoms~\cite{Kawaguchi12}. In the presence of a uniform external magnetic field, two ground-state phases of the spin-1 BEC support spin superflow in their order parameter manifold: the broken-axisymmetric phase (BA) for ferromagnetic interactions and the easy-plane polar (EPP) phase for antiferromagnetic interactions, where the BEC preserves spin rotation symmetry in the plane perpendicular to the magnetic field and a spin superflow for axial magnetization can be manifested with a spin texture. The stability of helical spin textures in the BA phase has been experimentally investigated~\cite{Vengalattore08}, and it was found that a uniform spin flow spontaneously decays into an irregular pattern due to dipolar interactions. For the EPP phase, collective spin-dipole oscillations were recently observed in a trapped BEC~\cite{Bienaime16}, demonstrating the existence of a stable spin superflow.

In this paper, we investigate the critical spin superflow dynamics of a spin-1 antiferromagnetic BEC by examining the damping of spin-dipole oscillations for large oscillation amplitudes. We observe two critical phenomena: generation of dark-bright solitons, which is explained by the modulation instability of the counterflow of two spin components~\cite{Hamner11,Law01}, and generation of transverse magnon excitations, which occurs via spin-exchanging collisions and leads to the transient formation of axial polar spin domains. We also observe the emergence of spin turbulence as the solitons decay and the axial polar spin domains relax into the EPP phase~\cite{Takeuchi10,Ishino11,Fujimoto12}. The results of this work establish spin superfluidity in the spin-1 antiferromagnetic BEC system and provide a comprehensive picture of the dissipation mechanisms in its spin sector.

Our experiment starts with the preparation of a BEC of $\approx 3.0 \times10^{6}$ $^{23}$Na atoms in the $|F,m_{F}\rangle=|1,0\rangle$ hyperfine state. The condensate is confined in an optical dipole trap with trapping frequencies of $\{\omega_{x},\omega_{y},\omega_{z}\} = 2\pi\times\{2.7,3.9,372\}~$Hz and its Thomas-Fermi radii are $\{R_{x},R_{y},R_{z}\}\approx\{230,160,1.7\}~\mu$m. For the peak atomic density $n$ of the condensate, the density and spin healing lengths are given by $\xi_{n}=\hbar/\sqrt{2mc_{0} n}\approx 0.7~\mu$m and $\xi_{s}=\hbar/\sqrt{2mc_2 n}\approx 5.7~\mu$m, respectively, where $m$ is the atomic mass and $c_{0,2}$ denotes the density (spin) interaction coefficients~\cite{Black07}. Because $\xi_s > R_z$ and $\xi_s\ll R_{x,y}$, the spin dynamics of the condensate is effectively two-dimensional. We prepare the condensate first in the $|m_z=0\rangle$ state in an external magnetic field of $B_z=30$~mG and rotate the spin direction from the $z$-axis to the $xy$ plane by applying a radio-frequency (rf) pulse. Then, we stabilize the EPP phase by immediately turning on a microwave field to tune the quadratic Zeeman energy to $q/h\approx-5.0~$Hz~\cite{Gerbier06,Bookjans11,Zhao14}.

A spin superflow is generated by applying a linear magnetic field gradient $B_q=\frac{\partial |B_z|}{\partial x}$ along the $x$ direction, which gives the spin a spatially varying Lamor frequency, inducing a coplanar spin texture. $B_{q}$ is turned on before application of the rf pulse and kept constant~\cite{add2}. Together with the trapping potential, the effect of the magnetic field gradient is the shifting of the trap centers for the $m_z=\pm 1$ spin components to opposite directions, which drives a spin-dipole motion of the trapped condensate~\cite{Bienaime16}. $B_q$ is calibrated using the initial rate of increase of the spatial frequency of Ramsey interference fringes between the $m_z=\pm 1$ spin components~\cite{Sadgrove13}.

Figure 1 shows the magnetization distributions of the condensate at various hold times $t$ after the rf pulse is applied for $B_q\approx0.6$~mG/cm and 1.3~mG/cm, which are measured via spin-sensitive phase-contrast imaging~\cite{Seo15,Shin06}. As $t$ increases, magnetizations of opposite signs accumulate in each end of the condensate, indicating the generation of spin flow in the condensate. For low $B_q$, the magnetization profile is restored back to the initial neutral flat profile after one oscillation period $T$ [Fig.~1(c)] and continues to oscillate. For high $B_q$, on the other hand, we observe that magnetization ripples, the length scale of which is on the order of a few $\xi_s$, are generated as the two spin components separate from each other, and the magnetization profile cannot be fully restored to the initial state at $t=T$, exhibiting spatial fluctuations [Fig.~1(g)]. This shows that spin-dipole oscillations are damped at high spin flow velocity via the generation of collective spin excitations.

\begin{figure}
\includegraphics[width=8.5cm]{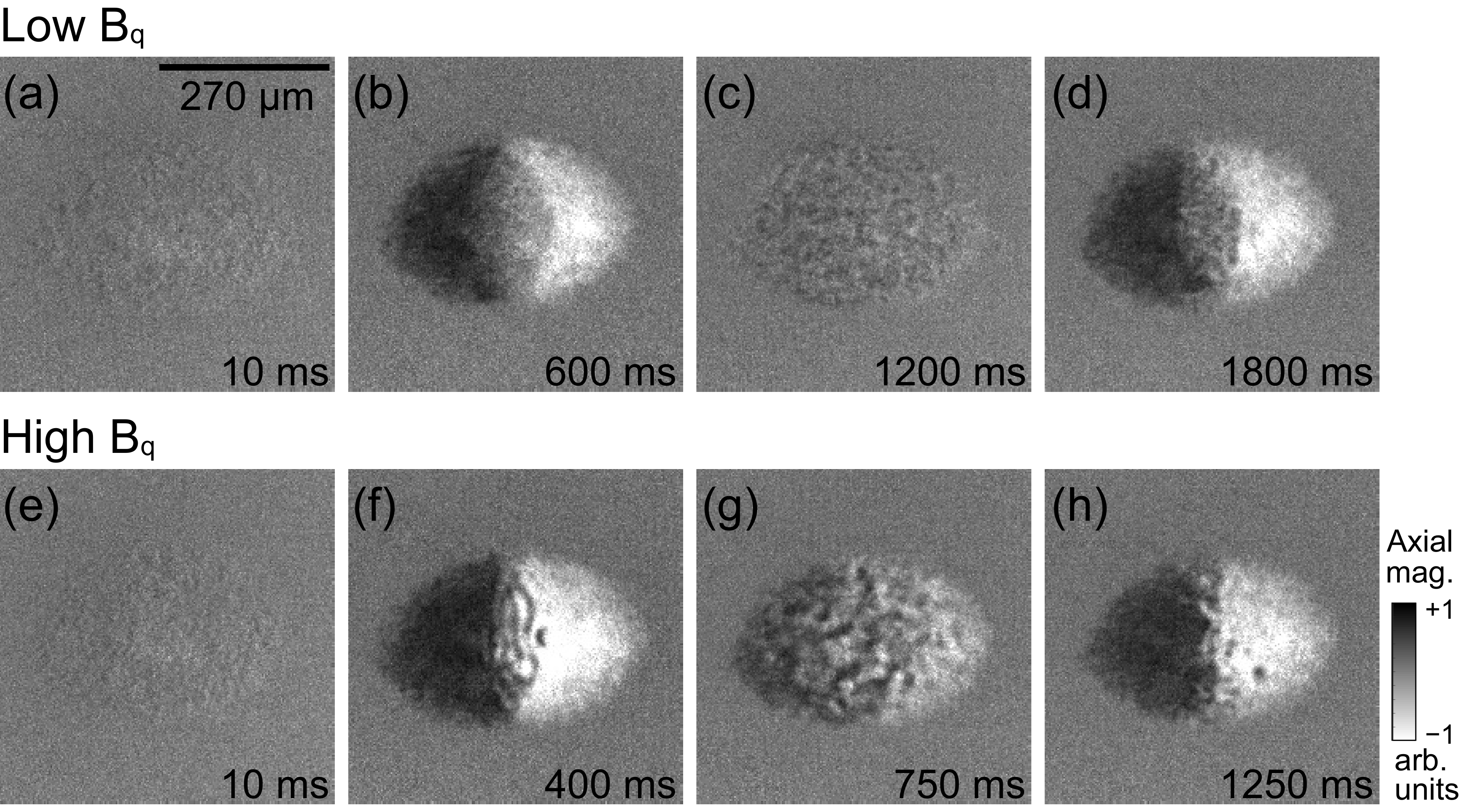}
\caption{Spin-dipole oscillations in a trapped antiferromagnetic spinor Bose-Einstein condensate. Spatial distribution of axial magnetization of the condensate at various hold times $t$ (a)-(d) for magnetic field gradient $B_q\approx0.6~$mG/cm and (e)-(h) 1.3~mG/cm. The oscillation period is $T\approx1230$~ms (840~ms) for $B_q\approx0.6~$mG/cm (1.3~mG/cm). The magnetization images were taken after 24$~$ms time-of-flight~\cite{Seo15}.}
\end{figure}

To quantify the damping of spin-dipole oscillations, we measure the time evolution of the separation $D(t)$ along the $x$-axis between the center-of-mass positions of the two spin components by employing a Stern-Gerlach (SG) spin-separation imaging technique~\cite{Bienaime16}. Defining the spin-dipole length $d_s \equiv \frac{2}{N} \int  x [n_+(\vec{r}) - n_-(\vec{r})] d^3r$, where $N$ is the total number of atoms and $n_\pm$ is the density distribution of the $m_z=\pm1$ spin component, $D$ is understood based on the relation $\Delta D(t)=D(t)-D(0)=d_s(t)+v_s(t)\tau$, where $v_s=\dot{d}_s$ and $\tau$ is the time-of-flight duration. We model damped sinusoidal oscillations of the spin-dipole length as $d_s(t) = A - Ae^{-\gamma t}(\cos\omega t + \frac{\gamma}{\omega}\sin\omega t)$, with amplitude $A$, frequency $\omega$, and damping rate $\gamma$, which satisfy the initial conditions of $d_s=0$ and $v_s=0$ at $t=0$. Then, $\Delta D(t) = A[ 1 - e^{-\gamma t}\{\cos(\omega t+\phi) /\cos \phi\} ]$, where $\tan \phi = - \frac{\gamma}{\omega} + \omega \tau + \frac{\gamma^{2}}{\omega}\tau$. We determine the parameters $\{A, \omega, \gamma\}$ based on the fitting to the data of $\Delta D(t)$. 

\begin{figure}
\includegraphics[width=8.0cm]{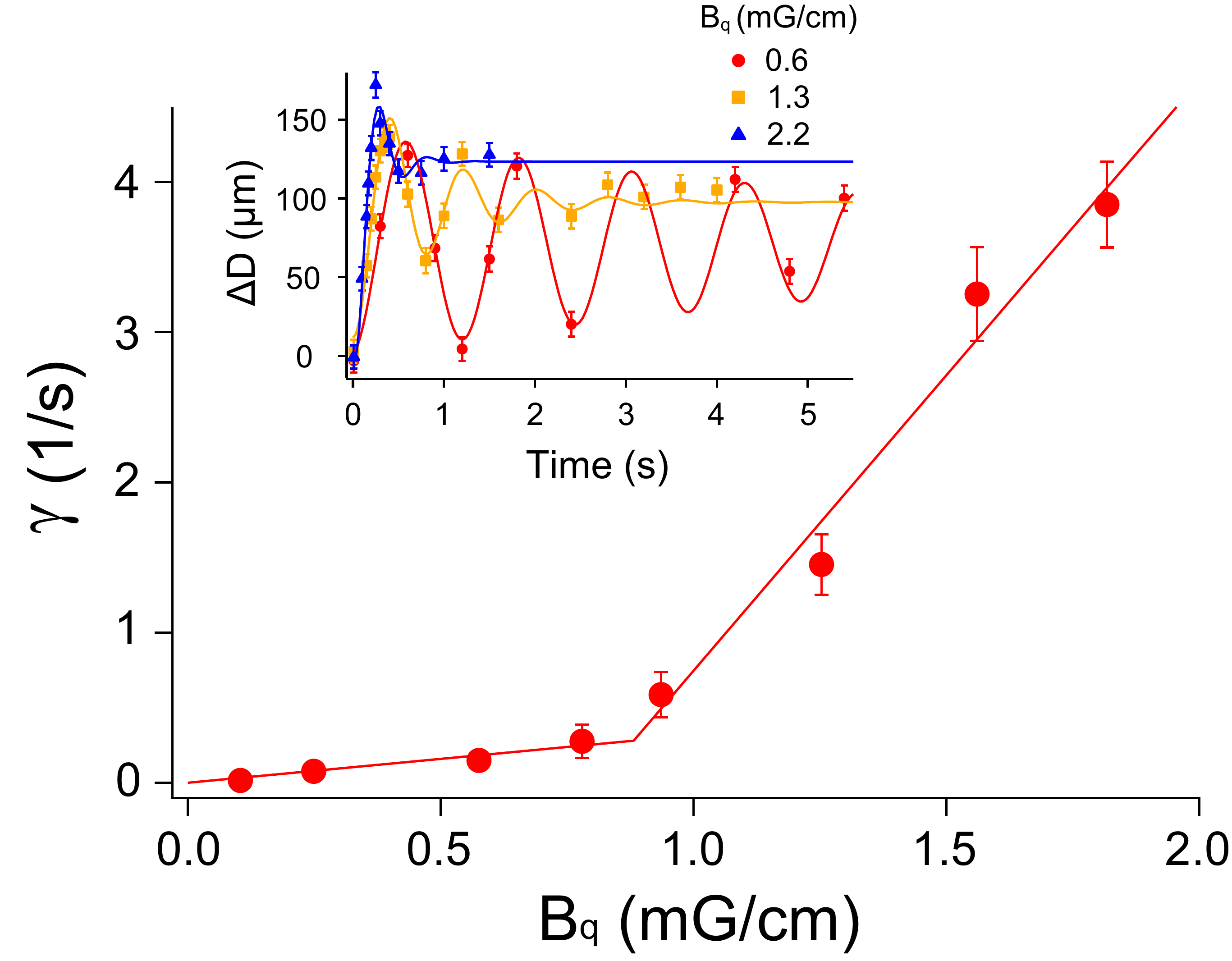}
\caption{Damping rate $\gamma$ as a function of the magnetic field gradient $B_q$. $\gamma$ was determined based on a damped sinusoidal function fit to the time evolution data of the relative displacement $\Delta D$ of the center-of-mass positions of the $m_z=\pm1$ spin components (inset); the error bar indicates the fitting error. The dashed and solid lines denote a two-line fit to the $\gamma$ data.}
\end{figure}

The damping rate $\gamma$ is observed to increase rapidly as $B_{q}$ increases over a certain value (Fig.~2). A two-line fitting to the measurement data gives a critical point at $B_{q}^{c}\approx0.9~$mG/cm. In particular, we verify that the critical point is identical to the minimum $B_q$, for which we can observe small-scale magnetic structures in the condensate at $t\approx \pi/\omega$ [Figs.~1(b) and 1(f)]. Demonstrating a finite critical velocity for spin superflow, this result establishes the spin superfluidity of the spinor superfluid system. In the region of small $B_q<B_q^c$, we observe a slight increase in $\gamma$ with increasing $B_q$, which we attribute to coupling to other low-lying spin excitation modes because of trap anharmonicity~\cite{Bienaime16,Sartori15}. The thermal fraction of the sample is less than $20\%$. At our lowest $B_q$, the spin-dipole oscillation frequency is measured to be $\omega/\omega_x\approx0.19$, which is consistent with the results in Ref.~\cite{Bienaime16}.

\begin{figure*}
\includegraphics[width=17.0cm]{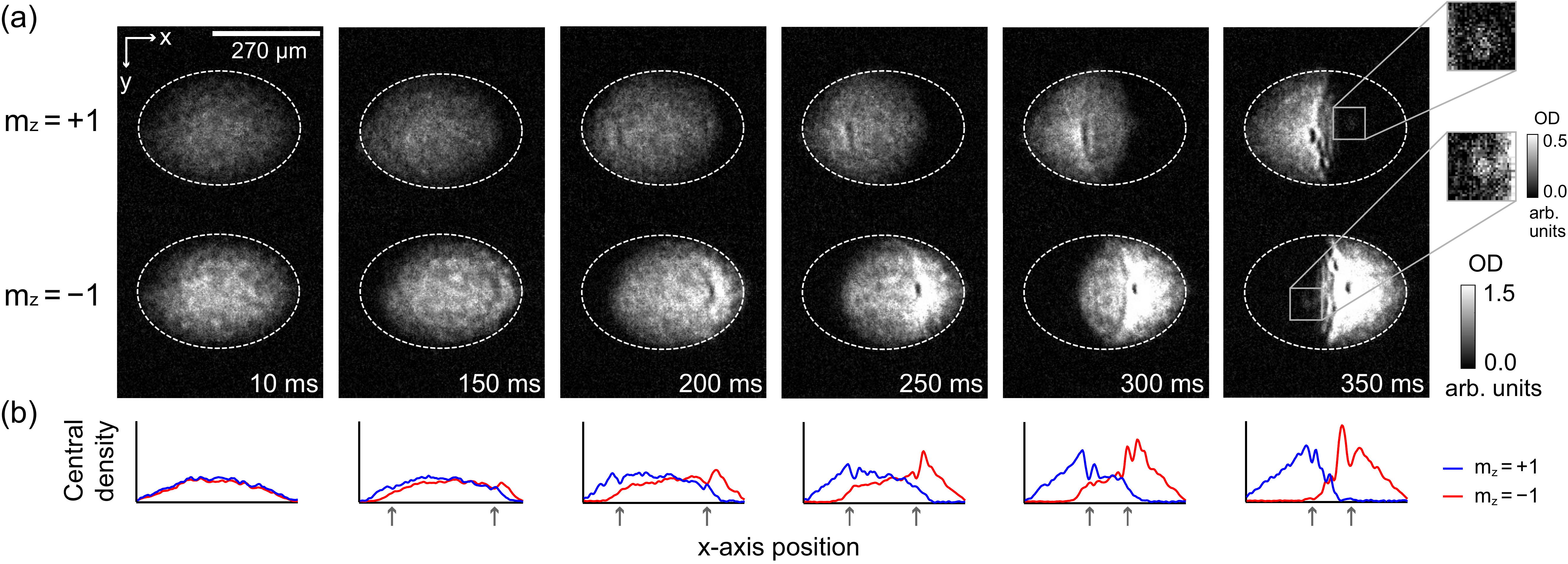}
\caption{Generation of dark-bright solitons. (a) Density distributions of the $m_z=\pm 1$ components for various hold times $t$ for $B_q\approx1.3$~mG/cm slightly above the critical point. The two spin components move to opposite directions, and dark-bright solitons are generated in the boundary of the two-component overlap region. The insets in the right column show enlarged images of the boxed regions, where the first dimple-shaped solitons are located. The dashed line indicates the condensate boundary. The optical depths (OD) of the two spin components are different because of the optical pumping effect that occurred during the imaging~\cite{Kim16}. (b) Density profiles of the two spin components along the $x$-axis in the central region. The gray arrows indicate the positions of the first solitons.}
\end{figure*}

In Fig.~3, to examine the generation dynamics of collective spin excitations, we display a sequence of the density distributions of the $m_z=\pm 1$ spin components for $B_{q}\approx1.3~$mG/cm slightly above the critical point. At $t\sim 150$~ms, density modulations along the spin flowing direction start appearing near the boundary of the overlap region of the two spin components. We verified that the condensate preserves a smooth density profile and that the density modulations of the two spin components are spatially anti-correlated. As the spin modulations are enhanced further, at $t\sim 250$~ms, a localized magnetization lump having a fully polarized core is formed at the tip of the overlap region. While more ripples are generated along the boundary of the overlap region, the magnetization lump propagates toward the center region, exhibiting its solitonic nature. This observation remarkably parallels the results of previous experiments with two-component BECs~\cite{Hamner11,Hoefer11}, where it was observed that dark-bright solitons are generated preferentially in the overlap interface of the two components above a critical counterflow velocity. Thus, we understand that the critical point observed for spin superflow arises from the modulation instability of the two counterflowing spin components~\cite{Law01}.

\begin{figure}[b]
\includegraphics[width=8.5cm]{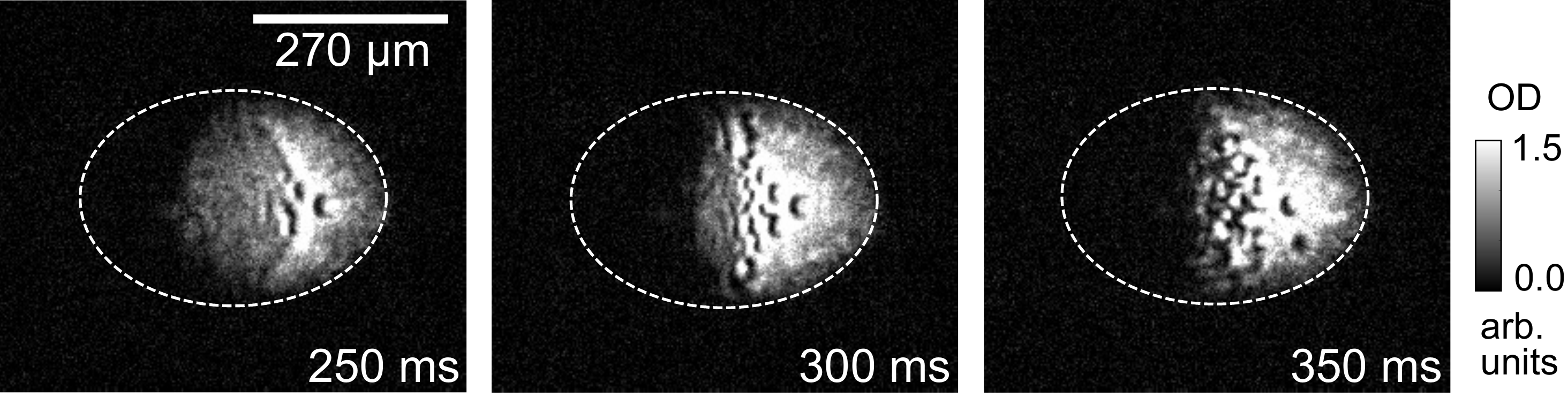}
\caption{Images of the $m_z=-1$ spin component at various hold times for  $B_{q}\approx1.6~$mG/cm. Solitons are progressively generated along the boundary of the overlap region. As the solitons decay into small segments due to their snake instability, a spin turbulence state forms in the center region.}
\end{figure}

From the analysis of the counterflow instability of a homogeneous binary superfluid~\cite{Abad15}, we find that the lower bound for the critical relative velocity is given by $v_{c1}=2\sqrt{c_2 \tilde{n}/m}$, where $\tilde{n}$ is the effective two-dimensional column density~\cite{Huang03}, and that $v_{c1}$ is not sensitive to spin polarization $p$. Taking into account the local density, we obtain $v_{c1}\approx 0.4$~mm/s for the region where the first soliton is generated. At $t\sim 200$~ms, i.e., the time at which the magnetization ripples develop, the propagation speed of the boundary of the overlap region is measured to be $v_{b}\approx 0.38~$mm/s, and the local spin polarization $p=(n_+-n_-)/(n_+ + n_-)$ around the first soliton formation position is $|p|\sim 0.3$. Assuming zero net mass flow, the local relative velocity between two spin components is estimated to be $v_{r}=2 v_{b}/(1+|p|)\sim 0.6~$mm/s, which is slightly higher than $v_{c1}$~\cite{add1}.  For a better quantitative comparison, it might be necessary to include the two-dimensional elliptical geometry of the condensate.

Figure 4 shows image data for higher $B_q\approx1.6$~mG/cm. After the first dimple-shaped soliton is generated at the tip of the overlap region, many dark-bright solitons are progressively generated along the boundary of the overlap region. In the two-dimensional geometry, the solitons decay into small segments via their snake instability~\cite{Feder00,Anderson01}. The length of the small segments is approximately 15~$\mu\textrm{m}\sim 3\xi_s$ in the center region of the condensate. Eventually, a turbulence state having a complex magnetization pattern is formed in the center region. The emergence of turbulence was theoretically anticipated in a counterflowing binary superfluid system~\cite{Takeuchi10,Ishino11}.

\begin{figure}[t]
\includegraphics[width=8.5cm]{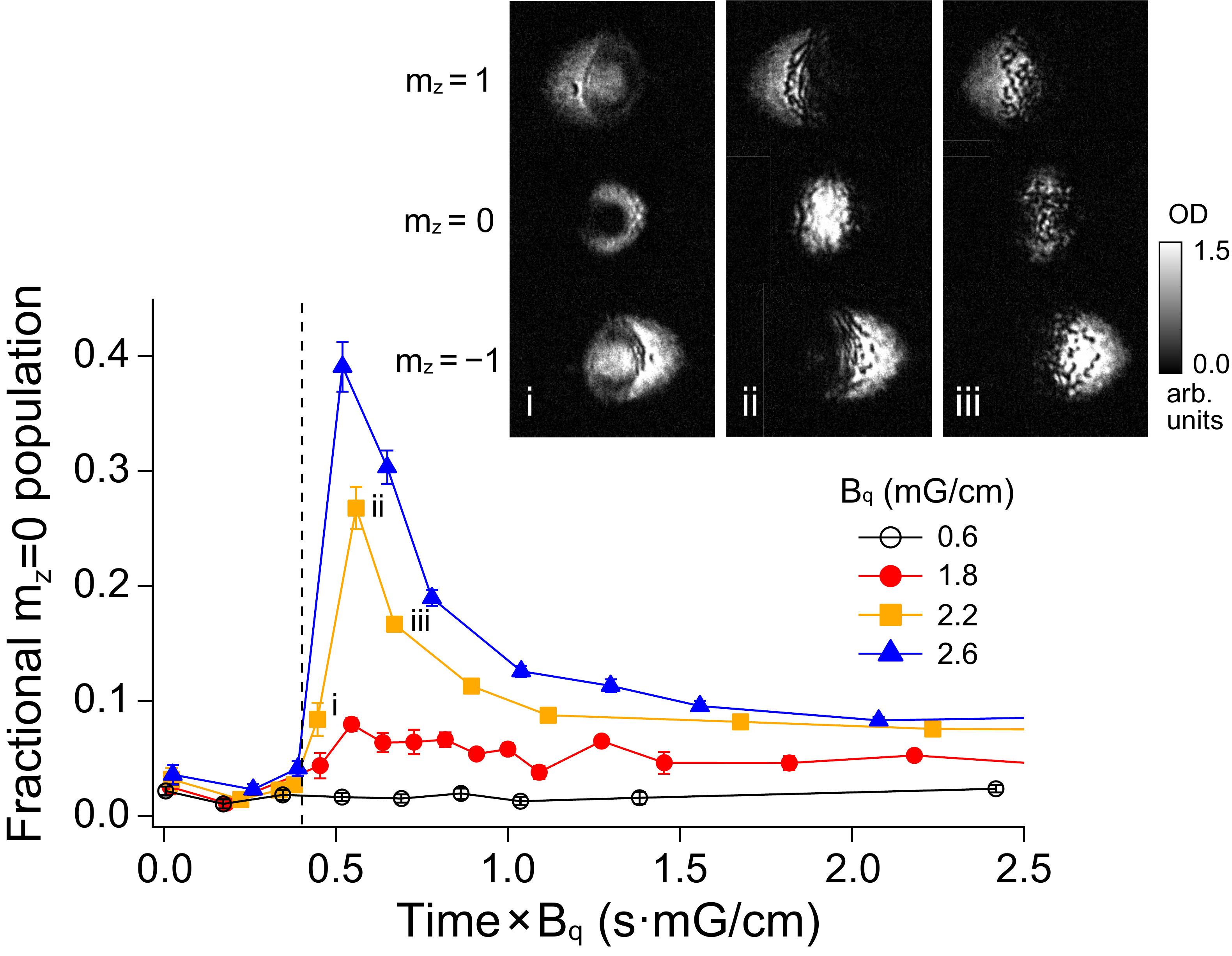}
\caption{Generation of transverse magnon excitations. Fractional population $\eta_0$ of the $m_z=0$ spin component as a function of the product of $t$ and $B_{q}$. When the spin flow velocity is high to overcome the quadratic Zeeman energy, spin-exchanging collisions occur to generate the $m_z=0$ spin components. The vertical dashed line indicates the onset point for $\eta_0$ at $t B_q \approx 0.4$~s$\cdot$mG/cm. The inset images show the density distributions of the spin components for $B_{q}\approx2.2~$mG/cm at (i) $t=200~$ms, (ii) $250~$ms and (iii) $300~$ms.}
\end{figure}

In our experiment, for high $B_q\geq 1.8$~mG/cm, we observe another critical phenomenon in which the $m_z=0$ spin component is created in a spin flow via spin-mixing collisions (Fig.~5). This process corresponds to the dynamic generation of transverse magnons, which are gapped excitations in a stationary EPP state. A condensate with a homogeneous spin current can be described as $\Psi=(\psi_{1},\psi_0, \psi_{-1})^{T}=\sqrt{\frac{n}{2}}(e^{ik_0 x}, 0, e^{-i k_0 x})^{T}$, where $\psi_{1,0,-1}$ denote the wave functions of the $m_z=1,0,-1$ spin components, respectively, and the spin flow velocity is $v_{r}=2 \hbar k_0 /m$. The Bogoliubov analysis of this spin-current-carrying state yields a transverse magnon mode, $\delta \Psi \propto (0, e^{ikx}, 0)^\textrm{T}$, with energy spectra of $E_k=\sqrt{(\epsilon_k + q')(\epsilon_k + q' +2 c_2 n)}$~\cite{Fujimoto12,Zhu15}, where $\epsilon_k=\hbar^2 k^2 /(2m)$ is the single-particle spectrum and $q'=|q| - \epsilon_{k_0}$. The gap energy is given by $\Delta_g=\sqrt{q'(q'+2c_2 n)}$ for $k=0$, and it decreases with increasing $\epsilon_{k_0}$. When $\epsilon_{k_0}>|q|$, i.e., $v_{r}> v_{c2}=\sqrt{8 |q|/m}$, $\Delta_g^2<0$, which means that the spin-current-carrying state becomes dynamically unstable and thus able to generate transverse magnons. It is observed that the $m_z=0$ component is initially created with a smooth spatial pattern in the center region of low $|p|$. This result is consistent with the Bogoliubov analysis results, in which the most unstable mode occurs at $k=0$ and the gap energy increases with increasing $|p|$.

In Fig.~5, the fractional population $\eta_0$ of the $m_z=0$ spin component is displayed as a function of $t B_q$. In all our measurements for high $B_q\geq 1.8$~mG/cm, $\eta_0$ shows sudden onset behavior at $t B_q \approx 0.4$~s$\cdot$mG/cm. Ignoring the effects of the trapping potential and atom interactions, the relative velocity of the $m_z=\pm 1$ spin components can be approximated to be $v_{r}=(\mu_B/m) t B_q$. Our measurement results suggest that $v_{r} \approx 0.9$~mm/s at the critical point, which is in good agreement with the prediction of $v_{c2}= 0.83$~mm/s for $|q|/h=5$~Hz. Under our experimental conditions, $v_{c1} < v_{c2}$ for the peak atom density. This means that the modulation instability of spin superflow is incurred in the center region before transverse magnon excitations are generated, and we infer that for a homogeneous system, the dynamic instability for generating magnon excitations is stronger than the modulation instability~\cite{Fujimoto12}. At the boundary of the overlap region, solitons are already created [Fig.~5 inset (i)].

The turbulence state involving the $m_z=0$ spin component is qualitatively different from that shown in Fig.~4 for lower $B_q$. Because of its immiscibility to the $m_z=\pm1$ components~\cite{Stenger_nat}, the generated $m_z=0$ component forms axial polar spin domains, wherein spin flow is completely absent. Then, the spin domains, having an excitation energy of $|q|$ with respect to the ground EPP state, dynamically relax with decreasing $m_z=0$ population, resulting in an irregular spin texture. The quench dynamics of the axial polar phase for negative $q$ were investigated in our recent work~\cite{Kang17}.

In conclusion, we have investigated the critical spin superflow dynamics of an antiferromagnetic spinor Bose-Einstein condensate and observed two critical phenomena: 1) dark-bright soliton generation due to the modulation instability of counterflowing spin components and 2) transverse magnon generation via spin-exchange collisions. An interesting extension of this work is to investigate the case of $|q|\rightarrow 0$, where a simple extrapolation of our result predicts $v_{c2}\rightarrow 0$~\cite{Fujimoto12}, i.e., the absence of spin superfluidity.

\begin{acknowledgments}
This work was supported by the Samsung Science and Technology Foundation under Project Number SSTF-BA1601-06.
\end{acknowledgments}


\begin{references}

\bibitem{Landau} L.~D.~Landau, The theory of superfluidity of helium II, J.~Phys.~USSR {\bf 5}, 71 (1941).

\bibitem{Sonin10} E.~B.~Sonin, Spin currents and spin superfluidity, Advances in Physics {\bf 59}, 181 (2010).

\bibitem{Jang11} J.~Jang, D.~G.~Ferguson, V.~Vakaryuk, R.~Budakian, S.~B.~Chung, P.~M.~Goldbart, and Y.~Maeno, Observation of half-height magnetization steps in 
Sr$_2$RuO$_4$, Science {\bf 331}, 186 (2011).

\bibitem{Manni12} F.~Manni, K.~G.~Lagoudakis, T.~C.~H.~Liew, R.~André, V.~Savona, and B.~Deveaud, Dissociation dynamics of singly charged vortices into half-quantum vortex pairs, Nat.~Commun. {\bf 3}, 1309 (2012).

\bibitem{Seo15} S.~W.~Seo, S.~Kang, W.~J.~Kwon, and Y.~Shin, Half-quantum Vortices in an Antiferromagnetic Spinor Bose-Einstein
Condensate, Phys.~Rev.~Lett. {\bf 115}, 015301 (2015).

\bibitem{Autti16} S. Autti, V. V. Dmitriev, J. T. Mäkinen, A. A. Soldatov, G. E. Volovik, A. N. Yudin, V. V. Zavjalov, and V. B. Eltsov, Observation of Half-Quantum Vortices in Topological Superfluid $^3$He, Phys.~Rev.~Lett. {\bf 117}, 255301 (2016). 


\bibitem{Bartolac81} T.~J.~Bartolac, C.~M.~Gould, and H.~M.~Bozler, Soliton Propagation in $^3$He-$A$, Phys.~Rev.~Lett. {\bf 46}, 126 (1981).

\bibitem{Choi12} J.~Choi, W.~J.~Kwon, and Y.~Shin, Observation of Topologically Stable 2D Skyrmions in an Antiferromagnetic Spinor Bose-Einstein Condensate, Phys.~Rev.~Lett. {\bf 108}, 035301 (2012).

\bibitem{Hivet12} R.~Hivet,	H.~Flayac,	D.~D.~Solnyshkov,	D.~Tanese,	T.~Boulier,	D.~Andreoli,	E.~Giacobino,	J.~Bloch,	A.~Bramati,	G.~Malpuech, and A.~Amo, Half-solitons in a polariton quantum fluid behave like magnetic monopoles, Nat.~Phys. {\bf 8}, 724 (2012). 


\bibitem{Choi13} J.~Choi, S.~Kang, S.~W.~Seo, W.~J.~Kwon, and Y.~Shin, Observation of a Geometric Hall Effect in a Spinor Bose-Einstein Condensate with a Skyrmion Spin Texture, Phys.~Rev.~Lett. {\bf 111}, 245301 (2013).

\bibitem{Seo16} S.~W.~Seo, W.~J.~Kwon, S.~Kang, and Y.~Shin, Collisional Dynamics of Half-Quantum Vortices in a Spinor Bose-Einstein Condensate, Phys.~Rev.~Lett. {\bf 116}, 185301 (2016).

\bibitem{Ollikainen17} T.~Ollikainen, K.~Tiurev, A.~Blinova, W.~Lee, D.~S.~Hall, and M.~M{\"o}tt{\"o}nen, Experimental Realization of a Dirac Monopole through the Decay of an Isolated Monopole, Phys.~Rev.~X {\bf 7}, 021023 (2017).


\bibitem{Zhu12} Q.~Zhu, C.~Zhang, and B.~Wu, Exotic superfluidity in spin-orbit coupled Bose-Einstein condensates, Europhys.~Lett. {\bf 100}, 50003 (2012).

\bibitem{Ozawa13} T.~Ozawa, L.~P.~Pitaevskii, and S.~Stringari, Supercurrent and dynamical instability of spin-orbit-coupled ultracold Bose gases, Phys.~Rev.~A {\bf 87}, 063610 (2013).


\bibitem{Borovik-Ramonov89} A.~S.~Borovik-Romanov, Yu~M.~Bunkov, V.~V.~Dmitriev, Yu~M.~Mukharskiy, and D.~A.~Sergatskov, Investigation of spin supercurrents in ${^3}B$, Phys.~Rev.~Lett. {\bf 62}, 1631 (1989).


\bibitem{Hamner11} C.~Hamner, J.~J.~Chang, P.~Engels, and M.~A.~Hoefer, Generation of Dark-Bright Soliton Trains in Superfluid-Superfluid Counterflow, Phys.~Rev.~Lett. {\bf 106}, 065302 (2011).

\bibitem{Hoefer11} M.~A.~Hoefer, J.~J.~Chang, C.~Hamner, and P.~Engels, Dark-dark solitons and modulational instability in miscible two-component Bose-Einstein condensates, Phys.~Rev.~A {\bf 84}, 041605(R) (2011).

\bibitem{Beattie13} S.~Beattie, S.~Moulder, R.~J.~Fletcher, and Z.~Hadzibabic, Persistent Currents in Spinor Condensates, Phys.~Rev.~Lett. {\bf 110}, 025301 (2013).

\bibitem{Kawaguchi12} Y.~Kawaguchi and M.~Ueda, Spinor Bose-Einstein condensates, Phys.~Rep. {\bf 520}, 253 (2012).


\bibitem{Vengalattore08} M.~Vengalattore, S.~R.~Leslie, J.~Guzman, and D.~M.~Stamper-Kurn, Spontaneously Modulated Spin Textures in a Dipolar Spinor Bose-Einstein Condensate, Phys.~Rev.~Lett. {\bf 100}, 170403 (2008).



\bibitem{Bienaime16} T.~Bienaim\'{e}, E.~Fava, G.~Colzi, C.~Mordini, S.~Serafini, C.~Qu, S.~Stringari, G.~Lamporesi, and G.~Ferrari, Spin-dipole oscillation and polarizability of a binary Bose-Einstein condensate near the miscible-immiscible phase transition, Phys.~Rev.~A {\bf 94}, 063652 (2016).




\bibitem{Law01} C.~K.~Law, C.~M.~Chan, P.~T.~Leung, and M.-C.~Chu, Critical velocity in a binary mixture of moving Bose condensates, Phys.~Rev.~A {\bf 63}, 063612 (2001).

\bibitem{Takeuchi10} H.~Takeuchi, S.~Ishino, and M.~Tsubota, Binary Quantum Turbulence Arising from Countersuperflow Instability in Two-Component Bose-Einstein Condensates, Phys. Rev. Lett. {\bf 105}, 205301 (2010).

\bibitem{Ishino11} S.~Ishino, M.~Tsubota, and H.~Takeuchi, Countersuperflow instability in miscible two-component Bose-Einstein condensates, Phys.~Rev.~A {\bf 83}, 063602 (2011).

\bibitem{Fujimoto12} K.Fujimoto and M.Tsubota, Counterflow instability and turbulence in a spin-1 spinor Bose-Einstein condensate, Phys.~Rev.~A {\bf 85}, 033642 (2012).

\bibitem{Black07} A.~T.~Black, E.~Gomez, L.~D.~Turner, S.~Jung, and P.~D.~Lett, Spinor Dynamics in an Antiferromagnetic Spin-1 Condensate, Phys.~Rev.~Lett. {\bf 99}, 070403 (2007).


\bibitem{Gerbier06} F.~Gerbier, A.~Widera, S.~F\H{o}lling, O.~Mandel, and I.~Bloch, Resonant control of spin dynamics in ultracold quantum gases by microwave dressing, Phys.~Rev.~A {\bf 73}, 041602(R) (2006).

\bibitem{Bookjans11} E.~M.~Bookjans, A.~Vinit, and C.~Raman, Quantum Phase Transition in an Antiferromagnetic Spinor Bose-Einstein Condensate, Phys.~Rev.~Lett. {\bf 107}, 195306 (2011).

\bibitem{Zhao14} L.~Zhao, J.~Jiang, T.~Tang, M.~Webb, and Y.~Liu, Dynamics in spinor condensates tuned by a microwave dressing field, Phys.~Rev.~A {\bf 89}, 023608 (2014).

\bibitem{add2} The variation of the Lamor frequency over the sample is tens of Hz, much smaller than the inverse of the rf pulse duration.


\bibitem{Sadgrove13} M.~Sadgrove, Y.~Eto, S.~Sekine, H.~Suzuki, and T.~Hirano, Ramsey Interferometry Using the Zeeman Sublevels in a Spin-2 Bose Gas, J.~Phys.~Soc.~Jpn. {\bf 82}, 094002 (2013).


\bibitem{Shin06} Y.~Shin, M.~W.~Zwierlein, C.~H.~Schunck, A.~Schirotzek, and W.~Ketterle, Observation of Phase Separation in a Strongly Interacting Imbalanced Fermi Gas, Phys.~Rev.~Lett. {\bf 97}, 030401 (2006).


\bibitem{Sartori15} A.~Sartori, J.~Marino, S.~Stringari, and A.~Recati, Spin-dipole oscillation and relaxation of coherently coupled Bose-Einstein condensates, New~J.~Phys. {\bf 17}, 093036 (2015).


\bibitem{Kim16} S.~Kim, S.~W.~Seo, H.-R~Noh, and Y.~Shin, Optical pumping effect in absorption imaging of F=1 atomic gases, Phys.~Rev.~A {\bf 94}, 023625 (2016).


\bibitem{Abad15} M.~Abad, A.~Recati, S.~Stringari, and F.~Chevy, Counter-flow instability of a quantum mixture of two superfluids, Eur.~Phys.~J.~D {\bf 69}, 126 (2015).

\bibitem{Huang03} G.~Huang, M.~G.~Velarde, and V.~A.~Makarov, Two-dimensional solitons in Bose-Einstein condensates with a disk-shaped trap, Phys.~Rev.~A {\bf 67}, 023604 (2003).



\bibitem{add1} We note that the results of Ref.~\cite{Knoop11} give a higher $c_2$ value than that from Ref.~\cite{Black07}, suggesting higher $v_{c1}\approx 0.6$~mm/s.

\bibitem{Knoop11} S.~Knoop, T.~Schuster, R.~Scelle, A.~Trautmann, J.~Appmeier, and M.~K.~Oberthaler, Feshbach spectroscopy and analysis of the interaction potentials of ultracold sodium, Phys.~Rev.~A {\bf 83}, 042704 (2011).




\bibitem{Feder00} D.~L.~Feder, M.~S.~Pindzola, L.~A.~Collins, B.~I.~Schneider, and C.~W.~Clark, Dark-soliton states of Bose-Einstein condensates in anisotropic traps, Phys.~Rev.~A {\bf 62}, 053606 (2000).

\bibitem{Anderson01} B.~P.~Anderson, P.~C.~Haljan, C.~A.~Regal, D.~L.~Feder, L.~A.~Collins, C.~W.~Clark, and E.~A.~Cornell, Watching Dark Solitons Decay into Vortex Rings in a Bose-Einstein Condensate, Phys.~Rev.~Lett. {\bf 86}, 2926 (2001).




\bibitem{Zhu15} Q.~Zhu, Q.-f.~Sun, and B.~Wu, Superfluidity of a pure spin current  in ultracold Bose gases, Phys.~Rev.~A {\bf 91}, 023633 (2015).



\bibitem{Stenger_nat} J.~Stenger, S.~Inouye, D.~M.~Stamper-Kurn, H.-J.~Miesner, A.~P.~Chikkatur, and W.~Ketterle, Spin domains in ground-state Bose-Einstein condensates, {Nature} \textbf{396}, 345 (1998).

\bibitem{Kang17} S.~Kang, S.~W.~Seo, J.~H.~Kim, Y.~Shin, Emergence and scaling of spin turbulence in quenched antiferromagnetic spinor Bose-Einstein condensates, Phys.~Rev.~A {\bf 95}, 053638 (2017).








\end{references}
\end{document}